\documentclass{ws-ijmpe} 
\usepackage{ulem} 
\begin{document} 
 
\markboth{A.G. Grunfeld, J. Berdermann, D.B. Blaschke, D. G\'omez Dumm,  T. 
Kl\"{a}hn, N.N. Scoccola}{Equation of state for hybrid compact 
stars with a nonlocal chiral quark model} 
 
\catchline{}{}{}{}{} 
%%%%%%%%%%%%%%%%%%%%%%%%%%%%%%%%%%%%%%%%%%%%%%%%%%%%%%%%%%%%%%%%%%%% 
 
\title{EQUATION OF STATE FOR HYBRID COMPACT STARS WITH A NONLOCAL CHIRAL QUARK 
MODEL} 
%\footnote{For the title, try not to 
%use more than 3 lines. Typeset the title in 10 pt Times roman, 
%uppercase and boldface.} 
 
\author{\footnotesize A.G. GRUNFELD} 
 
\address{Physics Department, CNEA, 
Av.\ Libertador 8250, 1429 Buenos Aires, Argentina \\ 
CONICET, Rivadavia 1917, 1033 Buenos Aires, Argentina \\ 
grunfeld@tandar.cnea.gov.ar} 
 
\author{J. BERDERMANN\footnote{Present address: DESY, Platanenallee 6, 
15738 Zeuthen, Germany}} 
 
\address{Institut f\"ur Physik, Universit\"at Rostock, 
Universit\"atsplatz 3, 18051 Rostock, Germany} 
 
\author{D.B. BLASCHKE} 
 
\address{Institute for Theoretical Physics, Univ.\ of Wroclaw, 
Max Born place 9, 50204 Wroclaw, Poland \\ 
Institut f\"ur Physik, Universit\"at Rostock, 
Universit\"atsplatz 3, 18051 Rostock, Germany \\ 
Bogoliubov  Lab.\ of Theor.\ Phys., JINR Dubna, 
Joliot-Curie Street 6, 141980  Dubna, Russia} 
  
\author{D. G\'OMEZ DUMM} 
 
\address{IFLP, Dpto.\ de F\'{\i}sica, Universidad Nacional de 
La Plata, C.C. 67, 1900 La Plata, Argentina\\ 
CONICET, Rivadavia 1917, 1033 Buenos Aires, Argentina} 

\author{T. KL\"AHN\footnote{Present address: Physics 
Division, Argonne National Laboratory, Argonne, IL 60439-4843, USA}} 
 
\address{Institut f\"ur Physik, Universit\"at Rostock, 
Universit\"atsplatz 3, 18051 Rostock, Germany} 
 
\author{N. N. SCOCCOLA} 
 
\address{Physics Department, CNEA, Av.\ Libertador 8250, 
1429 Buenos Aires, Argentina. \\ 
CONICET, Rivadavia 1917, 1033 Buenos Aires, Argentina.\\ 
Universidad Favaloro, Sol{\'\i}s 453, 1078 Buenos Aires, 
Argentina} 
 
\maketitle 
 
\begin{history} 
\received{(received date)} 
\revised{(revised date)} 
%\accepted{(Day Month Year)} 
%\comby{(xxxxxxxxxx)} 
\end{history} 
 
\begin{abstract} 
We study the thermodynamics of two flavor color superconducting (2SC) quark 
matter within a nonlocal chiral quark model, using both 
instantaneous and covariant nonlocal interactions.
For applications to compact stars, we impose conditions {of} electric and 
color charge neutrality as well as $\beta$ equilibrium and construct a 
phase transition to the hadronic matter phase described within the 
Dirac-Brueckner-Hartree-Fock (DBHF) approach.
We obtain mass-radius relations for hybrid star configurations which {fulfill} 
modern observational constraints, including {compact star} masses 
above $2M_\odot$. 
\end{abstract} 
 
\section{Theoretical framework} 
 
Although we understand hadrons as bound states of quarks, there is {no} 
unified approach available {yet} that adequately describe{s} the 
thermodynamics of the transition from nuclear to quark matter. Here we 
rely on a two-phase description, in which both states of matter are 
treated separately, using appropriate effective field theories. 
 
The hadronic matter EoS is obtained within the relativistic 
Dirac-Brueckner-Hartree-Fock (DBHF) theory\cite{honnef}. In this ab-initio 
approach, the nucleon inside the medium is dressed by a self-energy 
$\Sigma$, obtained from the Bethe-Salpeter equation for the 
nucleon-nucleon T-matrix in the ladder approximation using the Bonn-A 
potential in the interaction kernel. 
 
The quark matter phase is described within a nonlocal chiral quark model, 
which includes scalar and vector quark-antiquark interactions ---driven by 
coupling constants $G_S$ and $G_V$, respectively--- and anti-triplet 
scalar diquark interactions ---driven by a coupling constant $H$. We have 
carried out a systematic analysis considering both instantaneous and fully 
covariant nonlocal interactions, and study the corresponding parameter 
ranges. Explicit expressions for the nonlocal Lagrangians can be found in 
Refs.\cite{nosotros,Klahn:2006iw}.  
Our analysis is performed within bosonized versions of these quark models, 
in which scalar, vector and diquark bosonic fields are introduced. 
We consider the mean field approximation (MFA), expanding these fields around 
their respective vacuum expectation values (VEVs) and keep the lowest order 
relevant contributions (fluctuations are considered in the vacuum case to 
adjust model parameters from low energy phenomenology). The only 
nonvanishing mean field values in the scalar and vector sectors correspond 
to isospin zero fields $\bar\sigma$ and $\bar\omega$, respectively, while 
in the diquark sector, owing to the color symmetry, one can perform a 
rotation in color space ending up with a single nonvanishing VEV 
$\bar\Delta$. 
{I}n the presence of a baryochemical potential $\mu_B$, 
the thermodynamical potential per unit volume can be written as 
\begin{eqnarray} 
\Omega^{\mbox{\tiny MFA}} & = & \frac{ \bar \sigma^2 }{2 G_S} + \frac{\bar 
\Delta^2}{2 H} - \frac{\bar \omega^2}{2 G_V} - \frac{1}{2} \int 
\frac{d^4 \vec{p}}{(2\pi)^4} \, \ln \mbox{det} \left[\; 
S^{-1}(\bar \sigma,\bar \Delta,\bar 
\omega,\mu_{fc} ) \;\right] , \ \ \ \ \ \label{z} 
\end{eqnarray} 
where the inverse propagator $S^{-1}$ is a $48 \times 48$ matrix in Dirac, 
flavor, color and Nambu-Gorkov spaces (its explicit form for instantaneous 
and covariant models can be found in Refs. \cite{nosotros} and 
\cite{Klahn:2006iw} respectively). 
Since we require the quark matter to {be} electric and color charge neutral, 
we introduce different chemical potentials $\mu_{fc}$ for each quark flavor 
$f$ and color $c$. However, if the system is in 
chemical equilibrium all $\mu_{fc}$ can be written in terms of three 
independent quantities, namely the baryon chemical potential $\mu_B$, an 
electric chemical potential $\mu_Q$ and a color chemical potential 
$\mu_8$. 
 
{T}he values of $\bar\sigma$, $\bar\Delta$ and $\bar\omega$ can be 
obtained from {a} set of coupled gap equations 
\begin{equation} 
\frac{ d \Omega^{\mbox{\tiny MFA}}}{d\bar \Delta} \ = \ 0 \ , \hspace{1cm} 
\frac{ d \Omega^{\mbox{\tiny MFA}}}{d\bar \omega} \ = \ 0 \ , \hspace{1cm} 
\rule{0cm}{.7cm} \frac{ d \Omega^{\mbox{\tiny MFA}}}{d\bar \sigma} \ = \ 0 
\ . \label{sigud} 
\end{equation} 
Due to our nonlocal description, these VEVs come together with 
momentum-dependent form factors, which generate, e.g., momentum-dependent 
effective quark masses. 
 
In order to enable neutron stars to be in $\beta$-equilibrium,  
we have to account for electrons and muons. 
The latter can be treated as free Fermi gas contributions to the  
grand canonical thermodynamical potential. 
The electric and lepton chemical potentials turn out to be related by 
\begin{equation} 
\mu_{dc} - \mu_{uc} = - \mu_Q = \mu_e = \mu_\mu \ . 
\end{equation} 
Finally, $\mu_Q$ and $\mu_8$ become fixed by the conditions of 
vanishing electric and color densities. In this way, for each value of the 
temperature $T$ and chemical potential $\mu_B$ one can find the values of 
$\bar\Delta$, $\bar\sigma$, $\bar\omega$, $\mu_Q$ and $\mu_8$ that solve 
Eqs.~(\ref{sigud}). 
 
While the thermodynamics of the three-flavor case with color flavor locking 
(CFL) phases and selfconsistently determined mean fields could recently be 
achieved within the instantaneous chiral quark model \cite{Blaschke:2005uj},
the extension to the covariant case is a formidable task and deferred to 
future work.  
A justification of our restriction to the two flavor case in the present work 
is given by the following argument. 
In three flavor models the onset of strange quark matter is found at rather 
large densities. 
As a result hybrid star configurations with a strange matter core are found 
very close to the maximum mass of the corresponding two flavor configuration 
only\cite{Klahn:2006iw}. 
Since the corresponding color flavor locking phase in three flavor matter 
turned out to render compact stars unstable, 
the maximum compact star masses obtained in the present study might be 
slightly overestimated compared to a three flavor calculation 
but should be accurate within a range of $0.1~M_\odot$.  

\section{Numerical results and discussion} 
 
In order to obtain numerical predictions from our nonlocal models, we 
{set our} model parameters to be {in accordance} with $\mu=0$ phenomenology. 
In particular, we reproduce the empirical values of $m_\pi$ and $f_\pi$, and 
obtain a phenomenologically reasonable value of the chiral condensate $\langle 
0|\bar q q|0\rangle^{1/3} = -230$~MeV {by adjusting} the 
current quark mass, the scalar coupling $G_S$ and the model interaction 
range $\Lambda$. The couplings $G_V$ and $H$ as free parameters {have been 
arranged to ensure} accordance with phenomenological constraints from flow 
data analyses of heavy ion collisions\cite{Klahn:2006ir} and mass and 
mass-radius constraints from compact star observation{s}. 
Regarding the flow constraint, it is worth to mention that the hadronic 
DBHF EoS turns out to be too stiff for densities above $\sim 0.5$~fm$^{-3}$. 
This discrepancy is cured in our hybrid model 
after appropriately tuning the parameters: Since the hadron-to-quark matter 
transition softens the slope of the high density EoS, it becomes 
compatible with the data if the transition takes place at a critical 
density of $\sim 0.5$~fm$^{-3}$. 
 
Finally, {we demonstrate the compatibility of our models} with 
modern compact star observations. {In} Fig.\ 1 we 
show  sequences of compact star configurations obtained as solutions of 
the Tolman-\-Oppenheimer-\-Volkoff equations for selfgravitating hybrid star 
matter described by the theoretical schemes {presented above}.  
It can be seen that both covariant (dashed line in Fig.~1) and instantaneous 
(dash-dotted line) quark models are in good agreement with observational 
constraints. {In} particular, {they reproduce}  
the high mass of  pulsar PSR J0751$+$1807
and the mass-radius bounds from the isolated neutron star RX J1856 as well 
as the debated X-ray burster EXO 
0748$-$676~\cite{Ozel:2006bv,Alford:2006vz}.  
Thus, hybrid stars possessing quark matter cores appear to be compatible with 
all present observational constraints. 
 
\begin{figure} 
\hspace{2cm} 
\includegraphics[width=0.6\textwidth,angle=0]{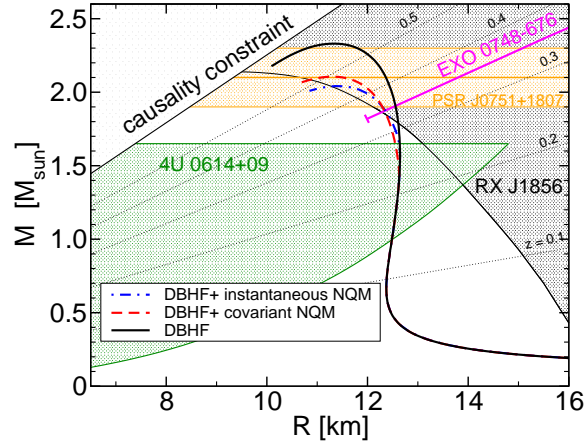} 
\caption{Mass-radius relations for neutron star configurations (DBHF EoS, 
solid line) and hybrid star configurations with hadronic shell and {a} color 
superconducting quark matter core for covariant (dashed) and instantaneous 
(dashed-dotted) nonlocal chiral quark models, see also 
Ref.~\protect\cite{Klahn:2006ir}.} 
\end{figure}

\section*{Acknowledgements} 
This work has been supported in part by CONICET and ANPCyT (Argentina), 
under grants PIP 6009, PIP 6084 and PICT04-03-25374, and by a scientist 
exchange program between Germany and Argentina funded jointly by DAAD and 
ANTORCHAS under grants No. DE/04/27956 and 4248-6, respectively. 
A.G.G. and D.G.D. thank for the hospitality of the IFT at University of 
Wroclaw, D.B.B. acknowledges support from the Polish Ministry of Science and 
Higher Education.

\end{document}